\documentclass[aps,prb,twocolumn,showpacs,preprintnumbers]{revtex4}

\usepackage{graphicx} 
\usepackage{tabularx}
\usepackage{dcolumn} 
\usepackage{bm} 
\usepackage{epsfig}
\usepackage{color}
\usepackage{multirow,longtable}
\usepackage{longtable}
\usepackage{amsmath}
\begin{document}


\title{Strong interactions, narrow bands, and dominant spin-orbit coupling  \\
in Mott insulating quadruple perovskite CaCo$_3$V$_4$O$_{12}$}

\author{H. B. Rhee and W. E. Pickett}
\affiliation{Department of Physics, University of California Davis, Davis CA 95616}

\date{\today}

\begin{abstract}
We investigate the electronic and magnetic structures and the character and direction of
spin and orbital moments of the recently synthesized 
quadruple perovskite compound CaCo$_3$V$_4$O$_{12}$ using a selection of methods from density functional theory. 
Implementing the generalized gradient approximation and the Hubbard $U$ correction (GGA+U), ferromagnetic spin 
alignment leads to half-metallicity rather than the observed narrow gap insulating behavior. Including spin-orbit 
coupling (SOC) leaves a half-semimetallic spectrum which is essentially Mott insulating.  SOC is crucial for the
Mott insulating character of the V $d^1$ ion, breaking the $d_{m=\pm 1}$ degeneracy and also giving a substantial
orbital moment. Evidence is obtained of the large orbital moments on Co that have been inferred from the measured 
susceptibility. Switching to the orbital polarization (OP) functional, GGA+OP+SOC also displays clear tendencies 
toward very large orbital moments but in its own distinctive manner.
In both approaches, application of SOC, which
requires specification of the direction of the spin, introduces large differences in the orbital moments of the 
three Co ions in the primitive cell. We study a fictitious but simpler cousin compound Ca$_3$CoV$_4$O$_{12}$ 
(Ca replacing two of the Co atoms) to probe in more transparent fashion the interplay of spin and orbital
degrees of freedom with the local environment of the planar CoO$_4$ units. The observation that the underlying 
mechanisms seems to be local to a CoO$_4$ plaquette, and that there is very strong coupling of the size of the 
orbital moment to the spin direction,
strongly suggest non-collinear spins, not only on Co but on the V sublattice as well.
\end{abstract}

\date{\today}
\maketitle

\section{Introduction}

Recently the Mott insulating quadruple perovskite CaCo$_3$V$_4$O$_{12}$ (CCVO) was reported by Ovsyannikov {\it et al.}\cite{ccvo}
having been synthesized under pressure. 
Several other quadruple perovskites with the formula $AA'_3B_4$O$_{12}$ have been studied, and this class of compounds with two
magnetic sublattices has been found to exhibit a wide range of intriguing phenomena, suggesting that its unusual structure may be playing 
a crucial role in the complex behavior that emerges. CaCu$_3$Mn$_4$O$_{12}$ (CCMO) gained attention for its colossal magnetoresistance 
(CMR),\cite{CCMO1999} and thus was compared to the manganates $Ln_{1-x}D_x$MnO$_3$ ($Ln=$lanthanide, $D=$alkali earth metal), 
which also exhibit CMR. Unlike the structurally simpler manganates, whose magnetoresistance (MR) becomes negligible at low applied 
magnetic fields, CCMO has a large low-field MR response spanning a wide temperature range below a high ferrimagnetic transition 
temperature $T_\mathrm c = 355$~K, a characteristic that is very attractive for magnetic sensor applications. CCMO was found to 
possess no Jahn-Teller Mn$^{3+}$ ions at the $B$ site and hence no double-exchange mechanism, which is what gives rise to CMR in 
the manganate perovskites. The origin of CMR (40\%) in CCMO, and very large low-field MR in BiCu$_3$Mn$_4$O$_{12}$ 
(Ref.~\onlinecite{BCMO}) and LaCu$_3$Mn$_4$O$_{12}$ (Ref.~\onlinecite{LCMO}) as well, appears to be due to spin tunneling or 
scattering at grain boundaries, rather than double exchange in the bulk. 

Another extreme property was found in the antiferromagnetic insulator CaCu$_3$Ti$_4$O$_{12}$, which displays a giant dielectric 
constant $\sim$$10^5$ that remains fairly constant from 100 to 500~K (Refs.~\onlinecite{Subra2000, Lin2002}). This dielectric
response was established to be extrinsic in nature, with the structure perhaps playing a central role in the defects that produce
very large polarization. Unusual magnetic 
transitions have been observed when gradually doping from ferromagnetic CaCu$_3$Ge$_4$O$_{12}$ to antiferromagnetic 
CaCu$_3$Ti$_4$O$_{12}$, and in turn to ferromagnetic CaCu$_3$Sn$_4$O$_{12}$, the mechanism of which is yet unknown but is 
apparently not due to bond angle-dependent superexchange interactions.\cite{CCGO-CCTO}

\begin{figure}[bt]
\begin{center}
\includegraphics[draft=false,width=0.9\columnwidth]{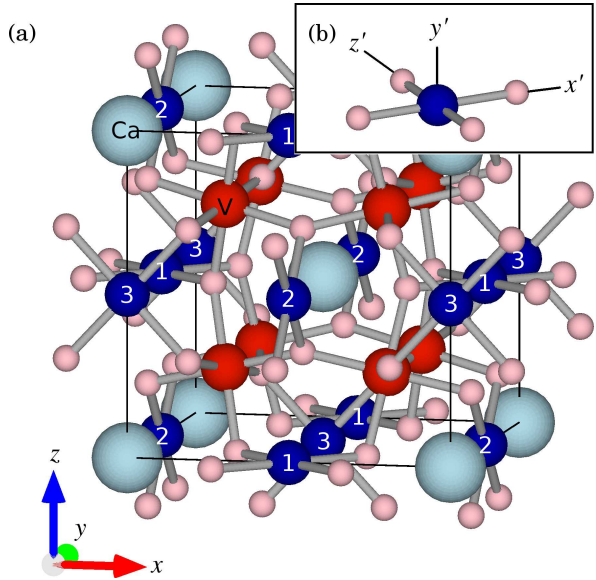}
\end{center}
\caption{(Color online) (a) Global coordinates and crystal structure of quadruple perovskite CCVO, containing two formula units per unit cell. The Co atoms (dark blue) are each surrounded by a plaquette of oxygens (pink), and the tilted VO$_6$ octahedra are nearly regular. The breaking of cubic symmetry due to spin-orbit coupling makes the three Co atoms (labeled 1 to 3) in the formula unit inequivalent. The magnetization quantization axis for our calculations, unless noted otherwise, is the $\hat z$ axis. To get the structure of fictitious compound Ca$_3$CoV$_4$O$_{12}$, replace Co atoms 1 and 2 with Ca. (b) Local coordinate system of a CoO$_4$ plaquette. Images were generated with VESTA.\cite{vesta}}
\label{struct}
\end{figure}

The structure of these materials provides important insight into, and restrictions on, the magnetic behavior, where a 
magnetic transition (apparently antiferromagnetic) is seen at $T_\mathrm{N} =98$~K.\cite{ccvo} Both the Co$^{2+}$ and V$^{4+}$ 
ions in this cobaltovanadate are expected to be magnetic and Mott insulating, and seemingly carries a substantial orbital moment to account for
the observed Curie-Weiss moment. The Co ion symmetry is $mmm$ in the rectangular CoO$_4$ plaquettes that promote,
as we will show, strong 
local anisotropy. Our calculations reveal that the Mott insulating character of the open-shell Co ion arises through the 
$d_{z^2}$ orbital in the local coordinate system. However, the overall cubic symmetry of the compound dictates that there are 
CoO$_4$ plaquettes perpendicular to each of the cubic axes, as can be seen in Fig.~\ref{struct}(a). The very narrow Co 
$3d$ bands -- very nearly localized orbitals -- will promote strong magnetic anisotropy, with spin moment either perpendicular to,
or within the plane of, the O$_4$ plaquette. 

The V$^{4+}$ $d^1$ ion, with its $\bar 3$ site symmetry, suggests a related conundrum. The four V ions have their individual symmetry 
axes along one of the four (111) directions. The occupied orbitals will be symmetry adapted linear combinations of the $t_{2g}$ orbitals. 
Crystal field considerations suggest the $a_{1g}$ symmetric combination will lie \textit{highest} in energy, leaving the doublet
usually denoted $e_g'$ singly occupied and therefore prime for orbital ordering. With the local (111) direction dictating the 
rotational symmetry, the $t_{2g}$ orbitals can be expressed in the form
\begin{eqnarray}
\psi_m = \frac{1}{\sqrt{3}}(\zeta_m^0 d_{xy} + \zeta_m^1 d_{yz} + \zeta_m^2 d_{zx}),
\label{eq:orb}
\end{eqnarray}
where $\zeta_m = e^{2\pi mi/3}$ is the associated phase factor for threefold rotations. There is vanishing orbital moment 
for the $a_{1g}$ member $m=0$, but potentially large orbital moments for the $e_g'$ pair $m=\pm 1$. 
These orbital moments, hence the spin and net moments, should tend to
lie along the (111) axis of symmetry of each V ion. Hence the V moments will be non-collinear among themselves, and
also not collinear with the Co moments. In such a situation, interpretation of the magnetization data 
in the ordered state will become a challenging problem.
We note that the non-collinearity that we are mentioning results from single-ion anisotropy rather than from
competing, frustrated exchange coupling, which is beyond the scope of the present study.

CCVO is currently the only quadruple perovskite reported that has Co exclusively occupying the $B$ site. 
Most known quadruple perovskites are half-metallic and semiconducting in the majority-spin channel; 
half-metallicity is often seen in perovskite oxides and other transition-metal oxides displaying 
CMR.\cite{LaBaMnO3-1997,Sr2FeMoO6-1998,LaMnCoO3-1997} In this paper, we perform density functional studies on CCVO, 
and our methods indicate how CCVO becomes a Mott insulator. We also study a related fictitious system, 
Ca$_3$CoV$_4$O$_{12}$, in which two of CCVO's three Co atoms are replaced by Ca, so that spin and orbital 
effects may be better understood on a local scale.



\section{Structure of CCVO and Calculation Methods}

CCVO takes on a structure that can be pictured as a variant of the cubic perovskite 
oxide \textit{AB}O$_3$. The superstructure $AA'_3B_4$O$_{12}$, with $Im\bar 3$ space group, is formed by quadrupling 
the parent unit cell and replacing $3/4$ of element 
$A$ with $A'$. Due to the introduction of $A'$, the symmetry of the structure is 
lowered by a large rotation of the $B$O$_6$ octahedra, which brings four oxygen ions 
closer to the $A'$ (Co) site to form a seemingly nearly square-planar environment, but we 
will show there is an important non-square component of the potential. Surprisingly, 
this particular quadruple perovskite houses VO$_6$ octahedra that are virtually regular: 
all V-O distances are identical, and the O-V-O angles deviate from $90^\circ$ by 
only $0.04^\circ$. The natural local axes of course do not align with the global crystal
axes. The CoO$_4$ plaquettes are not as regular, with the O-Co-O angles being 
$93.6^\circ$ and $86.4^\circ$. Because the non-square aspect becomes important, we will
refer to the unit as the CoO$_4$ plaquette.

Both Co and V ions are anticipated to be magnetic with Mott insulating character, so considerations of magnetic
coupling arise. V ions lie on a simple cubic sublattice separated by $a/2 = 3.67$~\AA, 
while Co ions lie on a bcc sublattice with the same nearest-neighbor Co-Co distance. 
The two perovskite $A$ and $B$ sublattices form a CsCl configuration, making it likely 
that nearest neighbor Co-V exchange interactions (versus Co-Co or V-V) 
are the driving force for magnetic order.

In oxides, Co and V often display strongly correlated behavior, so orbitally independent 
treatments such as GGA and local density approximation (LDA) do not provide the flexibility
to handle a compound like CCVO. We have therefore employed the GGA+U method,\cite{lsdau1,lsdau2} 
in which the intra-shell Coulomb repulsion $U$ and inter-orbital Hund's $J$ magnetic couplings were applied to
both Co and V with the following strengths: $U_\mathrm{Co} = 5$~eV, $J_\mathrm{Co} = 1$~eV, 
$U_\mathrm V = 3.4$~eV, $J_\mathrm V = 0.7$~eV.


All-electron calculations of CCVO and C3CVO were done with the {\sc WIEN2k}\cite{wien,wien2}
program, which is based on a full potential, linearized augmented planewave (FP-LAPW) method
within the density functional theory formalism. The Perdew-Burke-Ernzerhof flavor\cite{pbe}
of the GGA was chosen for the exchange-correlation functional. We used a
$17\times 17\times 17$ $k$-point mesh for the cubic unit cell, outlined in Fig.~\ref{struct}(a).
The sphere radii $R$ were set to 2.44, 2.04, 1.90, and 1.72~a.u.\ for Ca, Co, V, and O, respectively.
$RK_\mathrm{max} = 7.0$ was the cutoff for the planewave expansion of eigenstates. We used the CCVO
experimental lattice constant ($a=7.3428$~\AA) and internal coordinates,\cite{ccvo} and a collinear
ferromagnetic (FM) configuration was adopted for this first study, to obtain insight into the
intricate electronic and magnetic nature
of CCVO.

The orbital moment on Co, and thus the magnetocrystalline anisotropy in CCVO have been
suggested to be large,\cite{ccvo} reflecting the presence of strong spin-orbit coupling (SOC). It then is important to 
include SOC in the calculations.  Without SOC, the symmetry of CCVO is cubic (even with spin 
polarization) so each of the three Co atoms in the stoichiometric formula are equivalent. 
Taking into account SOC, with the direction of magnetization $\mathbf M$ in any one of the three axial directions, 
lowers the symmetry, and the three Co ions in their individually oriented CoO$_4$ plaquettes 
are no longer equivalent, and we find very large differences in the orbital and even the spin moments. 
In this paper 
the magnetization direction for all of our calculations will be along the $\hat z$ axis unless noted otherwise, 
and we will refer to the three different Co atoms as Co1, Co2, and Co3, labeled in Fig.~\ref{struct}(a)
as 1, 2, and 3, respectively. The respective CoO$_4$ plaquettes are perpendicular to the $\hat z$,
$\hat x$, and $\hat y$ directions respectively. 

\begin{figure*}[tb]
\begin{center}
\includegraphics[width=\textwidth]{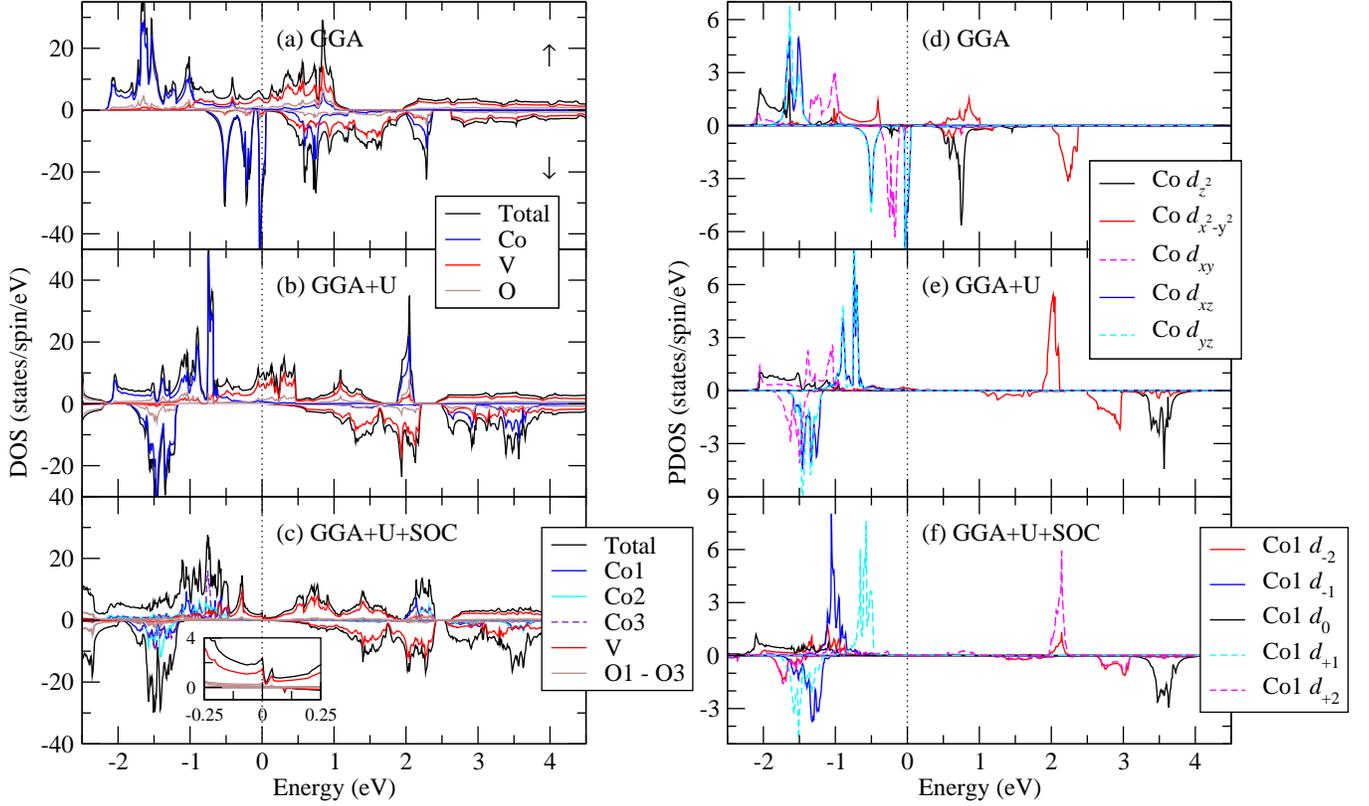}
\end{center}
\caption{(Color online) Spin-up and -down DOSs of CCVO resulting from FM alignment
for [left panels] (a) GGA, (b) GGA+U, and (c) GGA+U+SOC calculations, and 
[right panels] (d)--(f), the corresponding orbital- and spin-projected DOSs 
(in local coordinates) of Co. The top left (right) legend box applies to the two 
top left (right) graphs. The pseudogap mentioned in the text can be better seen in the blow-up of (c), inset in the bottom left panel. Ca does not contribute to the DOS within the featured range of energies. For the SOC-included calculations, the moment is along the $\hat z$ axis and the  projected DOS plot is for Co1.
}
\label{ccvo-dos}
\end{figure*}

Even with the addition of $U$, the orbital moment in Co perovskite oxides may be 
underestimated. In a fully relativistic treatment there is an additional orbital correction, 
referred to as orbital polarization (OP),\cite{brooks, eriksson} which is an attractive energy that is
proportional to the square of the orbital angular moment $\mathbf L$. This OP correction has
successfully been applied to intermetallics and especially to Co compounds that exhibit a large Co 
orbital moment.\cite{wu1992,trygg1995,sargol2006,wurmehl2006,cvo-saul2013,hollmann2014} The OP correction can be implemented in a phenomenological approach 
analogous to Hund's second rule that is implicit in GGA. Our results will show that 
orbital effects are indeed much larger within the OP scheme than within conventional GGA+U.

One one hand, the symmetry lowering due to SOC raises questions about the character of magnetic ordering that
occurs in CCVO. On the other hand, orbital moment formation and orientation seems to be a local
phenomenon, almost dictating non-collinear magnetic order.    
To simplify some of the local questions, we have studied a theoretical cousin  of CCVO, in which the Co2 and 
Co3 ions are replaced by simple Ca$^{2+}$, thereby isolating the Co1 ion for study. 
We present results for this model compound Ca$_3$CoV$_4$O$_{12}$, which we
refer to as C3CVO, to demonstrate that the local environment by and large governs the magnetic character of
the  CoO$_4$ unit in CCVO.


\section{Electronic structure analysis}
\subsection{CCVO}
The progression of the electronic and magnetic structure of FM CCVO with inclusion
of $U$ and then SOC is displayed in the various
panels of Fig.~\ref{ccvo-dos}. The right-hand panels allow identification of the most relevant orbitals---those of the partially
filled $3d$ shell of Co. When SOC is included, the angular momentum basis orbitals $d_m$ are used
for the projection, using the local coordinate system pictured in the Fig. 1 inset.

\vskip 3mm
\noindent\textit{Electronic structure within the GGA approach.} The collinear FM CCVO ground state within GGA
is metallic, as revealed in the density of states (DOS) in Fig.~\ref{ccvo-dos}(a). In all results to follow, occupation of the majority
Co states is subject to only one uncertainty: whether all majority $3d$ states are fully occupied, or
whether the $d_{x^2-y^2}$ orbitals, whose density lobes are directed toward the nearby O$^{2-}$ orbitals
and thereby form the highest lying crystal field orbital, are occupied (high spin [HS]) 
or unoccupied (low spin [LS]).
The minority projected DOS (PDOS)
shown in Fig.~\ref{ccvo-dos}(d) allows identification of the crystal field splittings of Co, since all
bands are narrow and peaks are evident. Relative to the GGA Fermi level, the crystal field picture 
of orbital energies is

\begin{center}
\begin{tabular}{ | l | r @{.} l | }
\hline
 $d_{x^2-y^2}$ &  2&25~eV\\
\hline
 $d_{z^2}$     &  0&75~eV\\
\hline
 $d_{yz}$      &  0&00~eV\\
\hline
 $d_{xy}$      & -0&20~eV\\
\hline
 $d_{xz}$      & -0&50~eV\\
\hline
\end{tabular}
\end{center}

It was previously suggested\cite{ccvo} based on the bond valence method that CCVO would have formal 
charges Ca$^{2+}$, Co$^{2+}$ $d^7$, V$^{4+}$ $d^1$, and O$^{2-}$. Our results are consistent with
this assignment, which also is the only reasonable assignment consistent with integral values
of formal charge.
Within GGA, the spectral density of the majority $d_{x^2-y^2}$ orbital (see Fig.~\ref{ccvo-dos}(d)) is
split above and below $\varepsilon_\mathrm F$---i.e., an itinerant band. 
The unfilled majority states' contribution to the spin moment is
drastically  reduced from the HS value of 3~$\mu_B$ to 1.77$\mu_B$, but far too large for LS.  

The majority V $t_{2g}$ bands extend from just below $\varepsilon_\mathrm F$ to 1~eV. The V spin of 0.77~$\mu_B$ 
is consistent with a $d^1$ configuration; however, partial filling of the essentially
localized minority Co $d_{xz}$ state
suggests some charge is taken from V. The very narrow Co minority $d_{yz}$, $d_{xy}$
states are filled. The localized $d_{yz}$ state with very high density of 
states at $\varepsilon_\mathrm F$, $N(\varepsilon_\mathrm F)$, pins the Fermi level and
implies, e.g., charge density wave or lattice instabilities. However, correlation effects and
SOC have not been taken into account yet.

\begin{figure}[bt]
\begin{center}
\includegraphics[draft=false,width=\columnwidth]{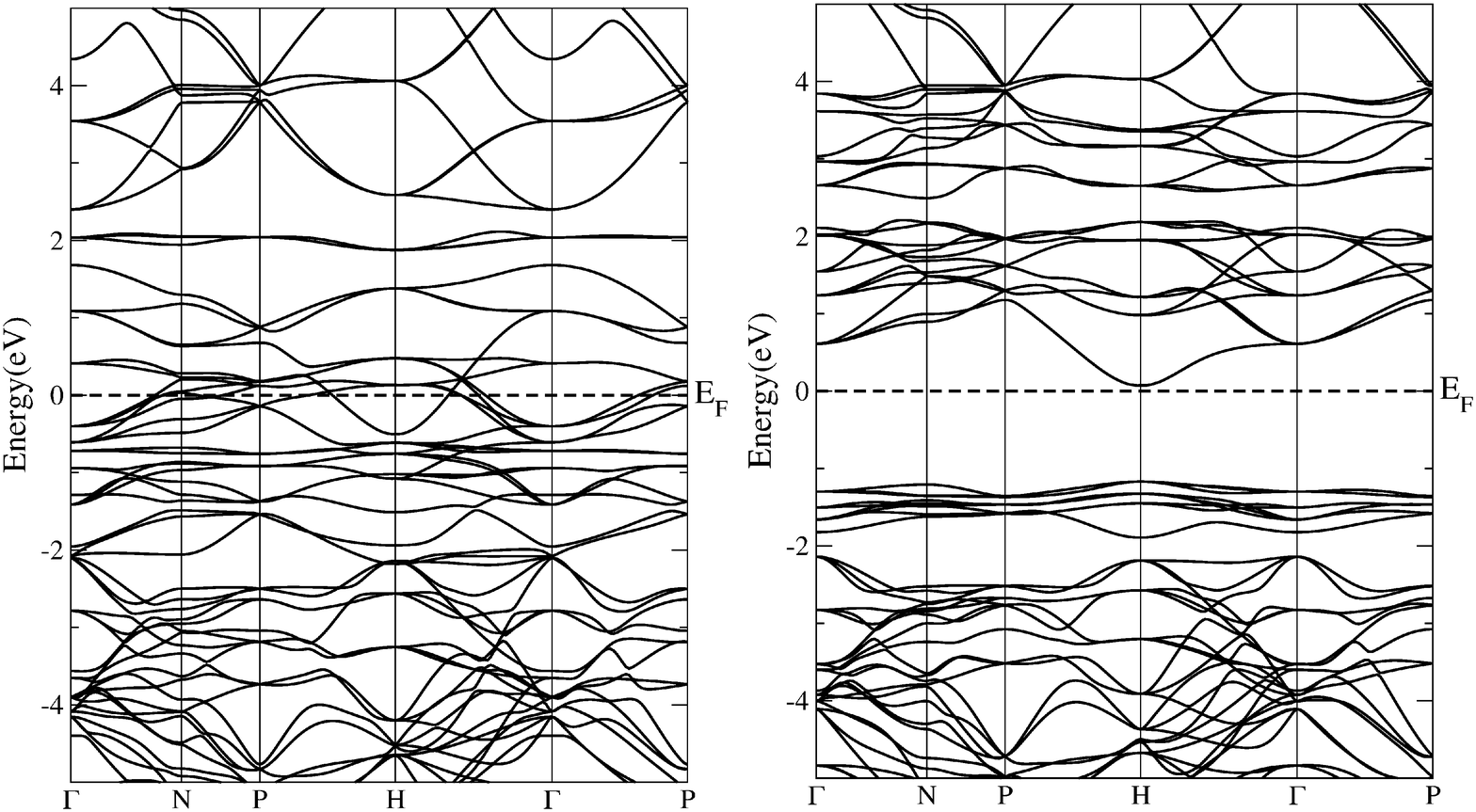}
\end{center}
\caption{Majority (left) and minority band structures resulting from a FM GGA+U calculation 
of CCVO. The lack of a Mott insulating gap is due to V bands, as discussed in the text.}
\label{bs-ggau}
\end{figure}

\vskip 3mm
\noindent\textit{Electronic structure within the GGA+U approach.} Adding an onsite Coulomb repulsion on each of the Co and V ions results in a half-metallic electronic structure. 
The minority gap is 1.25~eV, the change evident in Figs.~\ref{ccvo-dos}(a) and (b), and the corresponding GGA+U band structure displayed in Fig.~\ref{bs-ggau}. 
Calculated moments  are tabulated in Table~\ref{tbl:gga-moms}.
One effect of $U$ has been to remove
all Co states from near the Fermi level, leaving the clear Co$^{2+}$ formal charge 
with $d_{x^2-y^2}$ of both spins and minority $d_{z^2}$ unoccupied.
With four majority and three minority electrons on Co, $U$ has driven Co through a spin-state
transition, from roughly HS to clearly LS, with moment 1.0~$\mu_B$, as is common
for square-planar geometry.  
 The Co ion is Mott insulating in character.

Plotted in Fig.~\ref{ccvo-dos}(d) and (e) are the PDOSs of Co in CCVO from GGA and GGA+U calculations,
respectively. If the O$_4$ plaquette around Co were perfectly regular, the $d_{xz}$ and $d_{yz}$ states would be
degenerate. 
Applying $U_\mathrm{Co}$ lowers the sharp $d_{xz}$ and $d_{yz}$ peaks
in the minority channel, initially centered at $\varepsilon_\mathrm F$, to 1.5~eV below 
$\varepsilon_\mathrm F$. 
The $d_{z^2}$ orbital is affected most by $U_\mathrm{Co}$, the minority state being displaced higher
in energy by almost 3~eV. In the majority channel, most all of the states are shifted up; the formerly
divided $d_{x^2-y^2}$ states have coalesced to comprise the single hole in the majority channel.

Vanadium, on the other hand, drives the half-metallic nature by declining to open a
Mott gap, retaining a majority band crossing $\varepsilon_\mathrm F$ with $N(\varepsilon_\mathrm F) = 6.1$~eV$^{-1}$spin$^{-1}$.  
However, the symmetry used to this point retains the nearly cubic local symmetry of V and
more specifically the $e_g'$ degeneracy;
symmetry-breaking of some kind is required to break this symmetry and allow a Mott insulating
V configuration. However, SOC has not yet been taken into account and its effect is crucial. 

\begin{table}[bt]
\caption{Spin and orbital moments of the Co and V ions in CCVO, obtained from GGA, 
GGA+U and GGA+U+SOC calculations, as well as moments in C3CVO obtained from a GGA+U+SOC calculation. Moment are given in $\mu_\mathrm B$.}
\begin{ruledtabular}
\begin{tabular}{c|l|lrr} 
        &&& $\mu_\mathrm{spin}$ & $\mu_\mathrm{orb}$  \\
\hline
\multirow{3}{*}{CCVO}
&\multirow{2}{*}{GGA}
& Co        &  1.77  &   -     \\
&& V         &  0.79  &   -     \\ \cline{2-5}
&\multirow{2}{*}{GGA+U}
& Co1--Co3   &  1.00  &   -     \\
&& V         &  0.87  &   -     \\ \cline{2-5}
&\multirow{4}{*}{GGA+U+SOC}
& Co1       &  0.94  & $-0.75$    \\
&& Co2       &  0.93  & $-0.05$    \\
&& Co3       &  0.93  & $+0.07$    \\
&& V         &  0.84  & $-0.36$    \\
\hline
\multirow{1}{*}{C3CVO}
&\multirow{3}{*}{GGA+U+SOC}
& Co        &  0.97  & $-0.08$    \\
&& V         &  0.75  & $-0.07$    \\
\end{tabular}
\label{tbl:gga-moms}
\end{ruledtabular}
\end{table}

\vskip 3mm
\noindent\textit{{Electronic structure from GGA+U+SOC.}} Incorporating SOC lowers the
symmetry, one outcome being that the Co ions become inequivalent. Unexpected behavior 
was encountered at the GGA+SOC level when directing the spin along the (say) $\hat z$ axis: 
the three Co ions developed strong differences, violating the observed cubic symmetry
(although the differences might be difficult to see in standard x-ray diffraction).
Due to the very narrow Co $3d$ bands near or at the Fermi level, self-consistency became difficult
and the behavior was suggestive of multiple local minima with similar energies.

Related computational (mis)behavior arises even when lowering the symmetry by adding SOC \textit{after} 
GGA+U, when (our subsequent results show) all Co states are in the process of being removed 
from the Fermi level. Large orbitals moments arise,
which is surprising at first glance because $d_{+1}, d_{-1}$ or $d_{+2}, d_{-2}$ 
occupations (relative to the spin direction) must become unbalanced 
while their real and imaginary parts are nondegenerate and thus are not so readily mixed.
This choice of order in including $U$ and SOC results in antiferromagnetic (AFM)
tendencies---the spin of one of the three Co atoms reverses direction, and orbital
moments are vastly differing. 

To control the behavior of the Co moments, we devised the following procedure. Before adding $U$ or SOC, 
the symmetry was broken (for example, by displacing an atom by a physically negligible amount), which 
allowed the calculation to proceed from the GGA solution rather than to begin anew. In addition we enforced 
a total fixed spin moment (FSM) equal to that obtained from a calculation before the symmetry breaking; 
otherwise the artificial symmetry-breaking again makes self-consistency difficult. With the total moment 
fixed, the electronic and magnetic structures were relaxed. Then with the moment freed, $U$ was added 
and taken to self-consistency, 
and finally SOC was incorporated. The order of these last two steps was found to be important.
Carried out in this way, the unrealistically large differences between the Co ions was lessened. 
All results in this paper of GGA+U+SOC calculations for CCVO were obtained from this 
``infinitesimal broken symmetry + initial FSM'' protocol. We mention in passing that, since the 
experimental evidence suggested AFM order (at least not ferromagnetic order), we attempted
some AFM calculations, which require further doubling of the unit cell. These calculations 
encountered additional difficulties,
including resulting in considerably differing Co spins. We did not succeed in finding a procedure to 
avoid this in the AFM case,  and we discuss in the final section why doing so might 
not be any more realistic than
the FM ordered results that we present.

Including SOC dramatically changes the V-derived majority DOS at and around the Fermi energy, as 
shown in Fig.~\ref{ccvo-dos}(c). At the value of $U_\mathrm V$ that we have chosen (3.4~eV), 
the V manifold centered at $\varepsilon_\mathrm F$ opens, leaving a pseudogap. 
With our chosen values of $U$, this result leaves CCVO as a half-semimetal 
with a slight band overlap in the majority channel;
a somewhat larger value of $U_\mathrm V$ results in a FM Mott insulator state.
The minority spin spectrum, on the other hand, changes very little.  The gap is a result of 
relativity -- accounting for SOC -- which breaks the degeneracy of the V $d_{\pm 1}$ orbitals and
creates a substantial orbital moment of $-0.36 \mu_B$ that cancels 40\% of the spin moment.
Despite the significant changes to  the V ion when SOC is included, the formal valence remains unchanged.

Though the changes to the Co1 spectrum in Fig.~\ref{ccvo-dos}(f) seem less important,
they are such to drive an orbital moment of 0.75~$\mu_B$ on the Co1 ion. The unoccupied minority 
$d_{z^2}$ orbital becomes the $d_{0}$ peak and contributes nothing to the orbital moment.
The unoccupied majority $d_{x^2-y^2}$ orbital becomes about 85\% $d_{+2}$ and
15\% $d_{-2}$, thus providing the net orbital moment. This effect is enhanced
(perhaps, one might say, enabled) by the non-square symmetry of the CoO$_4$ unit.

\vskip 3mm
\noindent\textit{Beyond GGA+U+SOC.}
A fully relativistic treatment of the electronic structure reveals an orbital polarization (OP)
energy and associated potential analogous to that of spin polarization.
The OP method developed by Brooks\cite{brooks} and by Eriksson {\it et al.}\cite{eriksson}
is an orbitally dependent correction to energy functionals such as GGA and LDA. 
Hund's second rule is not taken into account in these functionals, leading to an 
underestimation of the orbital moment of many $d$ metals.\cite{daalderop1991} 
OP adds to the energy functional a correlation energy functional modeled as
\begin{equation*}
E_\mathrm{OP} = -\frac{1}{2} \sum_\sigma B_\sigma L_\sigma^2,
\end{equation*}
which acts to energetically favor larger orbital moments, while the resulting potential
makes $m_{\ell}$-dependent shifts in bands. Here, $B = F_2 - 5F_4$ is the Racah parameter 
in terms of the Slater integrals $F_2$ and $F_4$, $L$ is the orbital moment
in units of $\mu_B$, and $\sigma$ denotes spin. For Co, $B = 0.079$~eV; for V, 0.071~eV. 
The spin dependence ($<1$\%) of these paramters is included but is unimportant.

The moments from the OP method (GGA+OP+SOC) are listed at the top of Table~\ref{tbl:OP-moms}.
The largest Co orbital moment in the OP scheme is $1.5~\mu_\mathrm B$, more than double the 
value obtained from the GGA+U+SOC calculation. However, this time it is Co3, whose O$_4$ 
plaquette is perpendicular to $\hat y$, that possesses the large moment; the spin and
orbital moments lie with the CoO$_4$ plane rather than perpendicular to it. Co2 also has a 
large orbital moment, $0.9~\mu_\mathrm B$, again lying with the plane of the CoO$_4$
unit. The three Co spin moments range from  
1.7 to 2.0~$\mu_\mathrm B$; these more nearly HS values are (almost) twice the size 
of the LS moments from the GGA+U+SOC method. Thus OP enhances orbital moments but,
unexpectedly, feeds back to enhance greatly the spin moments while complicating
the atomic configurations of the Co ions.

\begin{table}[tb]
\caption{Spin and orbital moments of Co1, Co2, Co3, and V in CCVO, and Co and V for each $\hat z$, $\hat y$, and $\hat x$ magnetization direction in fictitious C3CVO, all from GGA+OP+SOC calculations. Units are in $\mu_\mathrm B$ and magnetization directions are in parentheses next to the atom.}
\begin{ruledtabular}
\begin{tabular}{ll|lrr} 
         && $\mu_\mathrm{spin}$ & $\mu_\mathrm{orb}$\\
\hline
\multirow{3}{*}{CCVO}
& Co1 ($\hat z$) &  1.69  &  0.25  \\
& Co2 ($\hat z$) &  1.83  &  0.87  \\
& Co3 ($\hat z$) &  2.05  &  1.54  \\
& V   ($\hat z$) &  0.75  & $-0.02$  \\
\hline
\multirow{3}{*}{C3CVO}
& Co ($\hat z$) &  1.52  &  0.27  \\
& Co ($\hat y$) &  1.51  &  0.43  \\
& Co ($\hat x$) &  1.77  &  1.11  \\
& V  ($\hat x,\hat y,\hat z$) &  0.71--0.73  & $-0.01$  \\
\end{tabular}
\label{tbl:OP-moms}
\end{ruledtabular}
\end{table}

\begin{figure}[tb]
\begin{center}
\includegraphics[draft=false,width=\columnwidth]{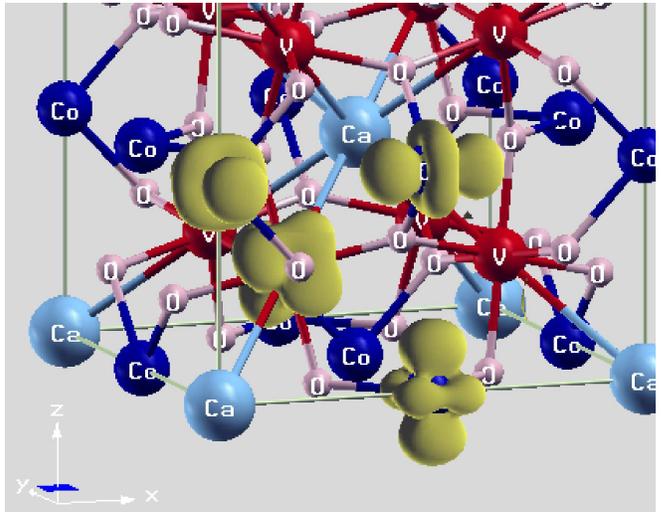}
\end{center}
\caption{(Color online) Spin-density isosurface of the valence state of CCVO obtained 
from a GGA+U+SOC calculation. The V isosurface is the central one. Differences of the
Co spin densities are discussed in the text.}
\label{sdens}
\end{figure}

The three-dimensional isosurface of the GGA+U+SOC valence spin density of CCVO is provided in 
Fig.~\ref{sdens}. The spin isosurfaces centered on each of the three cobalts take roughly a 
$d_{z^2}$-orbital shape, due to the Mott-Hubbard spitting of the Co $d_{z^2}$ orbitals. They do not 
align collinearly but along each O$_4$ plaquette's normal axis. The three show certain differences; 
the ``ring'' around the Co1 atom, when compared to that of Co2 and Co3, is more square-like, 
some $d_{x^2-y^2}$-like character in addition to the stronger $d_{z^2}$ character. 
This difference in spin density is a consequence of SOC -- without it, all three Co 
isosurfaces would be identical -- and its magnetization direction with respect to the 
direction of the O$_4$ plaquette.

The spin isosurface around V, shown also in Fig.~\ref{sdens}, is essentially identical
to its total valence \textit{charge} density isosurface, since the occupied V 3$d$ states originate 
from the majority channel only. These V $d$ states are concentrated between $-0.45$~eV  and the Fermi 
energy, and feature lobes corresponding to the linear combination in Eq.~\ref{eq:orb} and 
maintaining $\bar 3$ symmetry along the local (111) symmetry axis.

\vskip 3mm
\noindent\textit{Discussion of results for CCVO.} 
Inclusion of SOC, and more so with GGA+OP+SOC, indicates the likelihood of very large Co orbital
moments, as suggested by Ovsyannikov \textit{et. al} from analysis of the susceptibility.\cite{ccvo} Furthermore, the extreme
narrowness of the Co bands suggest the physics is local: the magnetic anisotropy is
determined by the CoO$_4$ configuration. The perpendicular axis (the local $\hat z$ axis)
seems the natural direction for a large orbital moment, and thus the spin as well.
While GGA+U+SOC gave that result, including OP shows that large orbital moments may well
lie within the plane of the CoO$_4$ unit. In any case, if the physics is local and each
unit behaves the same, the Co moments will be non-collinear. Similarly, the natural
axis for the V orbital moment is along its (111) symmetry direction, a different direction
for each V ion. 

The strong indications are that the magnetic order is non-collinear, possibly reinforced
by competing exchange interaction and suffering canting tendencies due to the 
Dzyaloshinskii-Moriya interaction between Co-V and V-V pairs. This combination of
strong correlation, important spin-orbit coupling in which the orbital moments feeds
back on the spin density, and intricate geometrical arrangement presents a daunting challenge
for an ab initio calculation. With substantial orbital moments, the magnetic coupling
between total moments becomes tensorial rather than scalar. Treating and understanding
the coupling would be a challenge even if the
exchange tensors were known, and it has recently been shown that obtaining
this tensor coupling when SOC coupling has large effects requires special 
technology.\cite{shu-ting}

CCVO's effective paramagnetic moment $\mu_\mathrm{eff}$ derived from magnetic 
susceptibility measurements\cite{ccvo} is 9.3~$\mu_\mathrm{B}$. Assuming the 
$S=\frac{1}{2}$ moment on V$^{4+}$, the effective net moment of Co is 4.4~$\mu_\mathrm{B}$, implying a considerable orbital moment of Co up to 
$\sim$2.5~$\mu_\mathrm B$, depending on the Co spin moment. Only recently have unusually 
large orbital moments for Co in perovskites been reported, yet not exceeding 1.8~$\mu_\mathrm{B}$. 
The possibility that V also has a moment does not help to account for the observed value,
since by Hund's rules it would cancel the spin moment, as shown in Table I. 

We do observe a large orbital moment of $0.7~\mu_\mathrm B$ on Co1 in the GGA+U+SOC 
result, but \textit{anti-parallel} to the spin. The other two cobalts have small,
typically sized orbital moments.  For the (001) spin direction, we have presumed, Co1 is
the distinctive Co, since its O$_4$ plaquette is the one that is perpendicular to the 
magnetization quantization axis, and its large moment lies along that axis. Orbital polarization, 
analogous to spin polarization (Hund's rule), has been found to improve calculated orbital 
moments in several magnetic materials, and it produces large orbital moments for CCVO. While 
$0.7~\mu_\mathrm B$ is an impressive value for a $3d$ orbital moment, OP more than doubles this to 
1.5~$\mu_\mathrm B$. The unique Co in the GGA+OP+SOC scheme is Co3 however, not Co1, such that the 
large orbital moment lies parallel to the O$_4$ plaquette.

\subsection{C3CVO} 

Continuing with the theme that the CoO$_4$ behavior involved primarily local physics, we
chose to adopt a simpler model that contains only one Co ion. Two of the Co$^{2+}$
ions are replaced with Ca$^{2+}$ ions, greatly simplifying both the conceptual issues
and the self-consistency process.  We proceed to investigate some of the behavior of
this model compound Ca$_3$CoV$_4$O$_{12}$ (C3CVO).

\begin{figure}[tbh]
\begin{center}
\includegraphics[draft=false,width=\columnwidth]{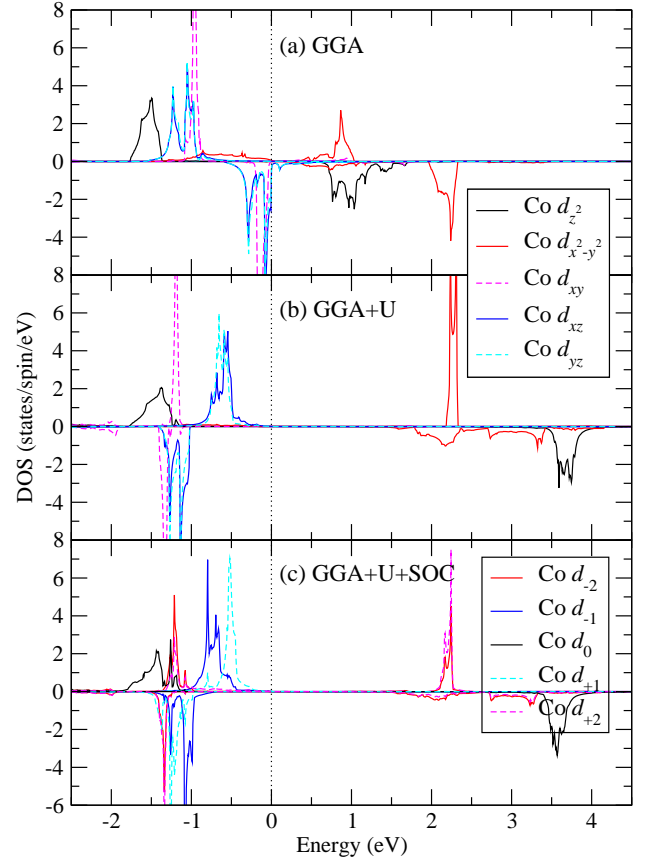}
\end{center}
\caption{(Color online) Spin-up and -down, orbitally projected PDOSs of Co in CCVO resulting from FM alignment using (a) GGA, (b) GGA+U, and (c) GGA+U+SOC(001) calculations. 
The top legend box applies to the top two graphs.}
\label{c3cvo-pdos}
\end{figure}

\vskip 3mm
\noindent\textit{GGA+U+SOC.} 
We have set the magnetization direction for C3CVO to be perpendicular 
to the O$_4$ plane on which the remaining Co lies; this is equivalent to keeping Co1, and replacing 
Co2 and Co3 with Ca in Fig.~\ref{struct}(a).
Comparing the PDOS in Fig.~\ref{c3cvo-pdos} for C3CVO to the right-hand panels of Fig.~\ref{ccvo-dos}
for CCVO, it is apparent that the electronic environment of Co in C3CVO is similar to that of Co1 in CCVO when SOC is included. This confirms that C3CVO represents reasonably the local planar 
environment of Co1 in CCVO without the complications from the other Co ions. The 
general characteristics that GGA, $U$, and SOC produced in C3CVO are very much like those 
seen in CCVO. With the GGA+U+SOC method, C3CVO, like CCVO, remains metallic in the majority channel, 
and semimetallic in the minority. A pseudogap, near Mott insulating, feature made up of V states at the Fermi energy
(not shown), similar to that in CCVO, is also present in C3CVO. 

Unexpectedly, the orbital moments of V and Co in C3CVO are 
much smaller that on the V and analogous Co1 in CCVO, as displayed at the bottom of Table~\ref{tbl:gga-moms}. 
In C3CVO, the orbital moment of V of 0.07~$\mu_\mathrm B$ is only one fifth of its value in CCVO. 
We note we have not oriented the spin along a (111) direction, where the V orbital moment may be larger.
The difference in the Co orbital moment is even larger, with C3CCVO's Co moment 
(0.08~$\mu_\mathrm B$) having only 10\% of the strength that it has in CCVO 
(0.75~$\mu_\mathrm B$), due perhaps to the absence of the Co-V network of exchange.

\vskip 3mm
\noindent\textit{GGA+OP+SOC.} Referring back to Fig.~\ref{struct}(a), when CCVO's quantization axis 
is the $\hat z$ axis, Co1 is analogous to the sole Co in C3CVO when \textit{its} quantization axis 
is also the $\hat z$ axis. Similarly, Co2 (Co3) is analogous to the only Co in C3CVO when quantization 
is in the $\hat y$ ($\hat x$) direction. The analogies are reflected by the ordering of the entries in 
Table~\ref{tbl:OP-moms}. The trends in CCVO and C3CVO are common, both in spin and orbital moments, 
which increase from Co1 to Co2 to Co3, and also as the magnetization axis is rotated in C3CVO from 
$\hat z$ to $\hat y$ to $\hat x$. In the GGA+U+SOC scheme, the difference between the orbital moment 
of Co1 in CCVO and that of Co in C3CVO was ten-fold. OP however lessens this difference: the largest 
orbital moment of Co in C3CVO is $1.1~\mu_\mathrm B$, which is 75\% of the maximum orbital moment 
of $1.5~\mu_\mathrm B$ in full structure of CCVO. This difference
arises from the differences in Co $3d$ occupation.

\begin{figure}[tb]
\begin{center}
\includegraphics[draft=false,width=\columnwidth]{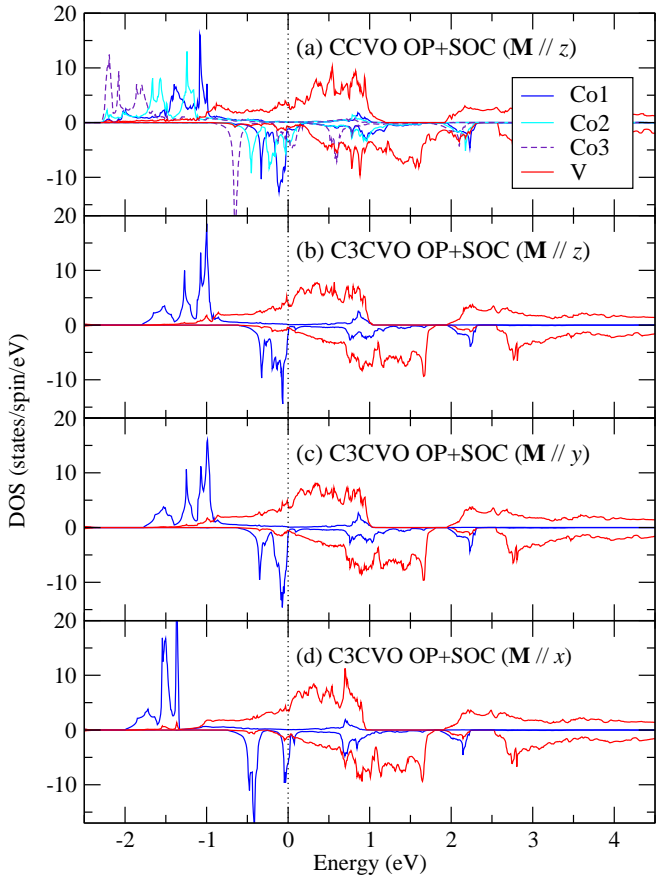}
\end{center}
\caption{(Color online) GGA+OP+SOC DOSs of cobalt(s) and V in (a) CCVO and (b)-(d) C3CVO. $\mathbf M$ is parallel to the (a)-(b) $\hat z$, (c) $\hat y$, and (d) $\hat x$ axes.}
\label{op-dos}
\end{figure}

Co in C3CVO has the largest orbital moment when $\mathbf M \parallel \hat x$, with an analogous result 
in CCVO---i.e., Co3 has the largest $\mu_\mathrm{orb}$. The uniqueness of Co3 in the OP+SOC method is 
curious, since it is the Co1 plaquette that is perpendicular to the quantization 
axis. The strong coupling of the size of $\mu_{orb}$ to the local spin direction can be understood by 
studying the Co DOSs presented in Fig.~\ref{op-dos}. In Fig.~\ref{op-dos}(a), the three Co DOSs are 
compared in one plot. Panels (b), (c), and (d) below it are the Co states of C3CVO with quantization 
in the various orientations, and it is clear that the DOSs are similar when $\mathbf M \parallel \hat z$ 
and $\mathbf M \parallel \hat y$ while different when $\mathbf M \parallel \hat x$. The spin-up 
occupied Co states are lower in energy when $\mathbf M \parallel \hat x$, and the down-spin occupied 
peak, instead of butting up against the edge of the Fermi level, has states that pass above 
$\varepsilon_\mathrm F$. These additional unoccupied states provide the distinction between 
$\mathbf M \parallel \hat x$ and $\mathbf M$ along the other two axial directions. 
The orbital potential evidently produces important shifts in positions of Co $3d$ bands.

In Fig.~\ref{op-dos} providing the PDOSs of CCVO and C2CVO for the OP+SOC results, one can observe 
that the Co1 electronic states in CCVO in panel (a) are very similar to 
those of C3CVO in (b) and (c). Meanwhile, Co3 of CCVO, like Co of C3CVO in Fig.~\ref{op-dos}(d) 
($\mathbf M \parallel \hat x$), has more strongly bound majority occupied states, and unoccupied 
states ``leak'' out of the minority peak centered just below $\varepsilon_\mathrm F$. Co2 in CCVO 
and Co in C3CVO when $\mathbf M \parallel \hat y$ do not share as many similarities as the other 
two analogous pairs, and that is reflected in the differing spin and orbital values in Table~\ref{tbl:OP-moms}. 

\begin{table}[tbh]
\caption{$E_\mathrm{OP}$ values of spin-down cobalts in CCVO and C3CVO, from OP+SOC calculations. Direction of $\mathbf M$ is in parentheses next to the atom.}
\begin{ruledtabular}
\centering
\begin{tabular}{l|lr} 
         && $E_\mathrm{orb}$ (eV) \\
\hline
\multirow{3}{*}{CCVO}
& Co1 ($\hat z$) &  0.01  \\
& Co2 ($\hat z$) &  0.06  \\
& Co3 ($\hat z$) &  0.19  \\
\hline
\multirow{3}{*}{C3CVO}
& Co ($\hat z$) & $-0.01$  \\
& Co ($\hat y$) & $-0.02$  \\
& Co ($\hat x$) &  0.10  \\
\end{tabular}
\label{tbl:OP-Eorb}
\end{ruledtabular}
\end{table}

\section{Discussion and Summary}

Our several investigations into the electronic and magnetic structure of CCVO reveal a
great deal of complexity, even considering the three Co and four V ions in the primitive
cell, each with its own natural axis of exact (V) or approximate (Co) symmetry. 
This complexity is the result of 
(1) two types of open shell cations, 
(2) strong correlation in narrow bands, 
(3) important and in some cases dominating spin-orbit coupling, and 
(4) an intricate three dimensional network.
To account for the insulating
nature, both Co and V must display Mott insulating character. While the crystal field
splittings on the Co site -- a span of 2.75 eV -- are typical for a $3d$ ion in an oxide, the
subsplittings become very important
given the very narrow Co $3d$ bands. 
Though these bands are narrow and separate above and below the Fermi level, the Co spin is not
very representative of either the HS nor LS state, a reflection of some charge transfer
character of this cobaltovanadate.
The O$_4$ plaquette has an
important non-square components of the potential, so spin directed in the 
$\hat x$ and $\hat y$ directions produce
very different orbital moments. 

We have not given to the V ion the attention that it may deserve. Each of the V ions
has its own threefold axis (symmetry related, of course), which provides a natural
axis for the orbital moment, and hence the spin moment. The $d^1$ configuration in the
$e_g'$ doublet requires SOC, orbital ordering, or structural distortion to provide
the  splitting necessary for a Mott gap. 

In addition to the widely applied GGA+U+SOC method, we have explored the application of
the little used orbital polarization potential via GGA+OP+SOC. This OP is
an analog of the spin polarization potential within GGA but is roughly an order of
 magnitude smaller, but it arises in a fully relativistic theory of
electronic structure and should be applied more widely and studied. 
Both methods give a tendency toward very large orbital moments,
but the behavior is different in the two approaches.

Whichever approach is used, only one of the three Co ions acquires a large $\mu_\mathrm{orb}$
when the spin moments (hence orbital moments) are restricted to be collinear. 
This observation, together with the
large orbital moment inferred from experiment and the fact that the mechanisms of
magnetization appear to be local, suggest that the Co moments will be non-collinear,
each directed along its own large $\mu_{orb}$ axis. Analogously, the V moments are
likely to be non-collinear as well, and in different directions [(111) axes] than the
Co moments.   

CaCo$_3$V$_4$O$_{12}$ thus presents a very challenging electronic and magnetic structure
problem. After starting from the GGA starting point, both the Hubbard $U$ and spin-orbit
coupling---together with the required great reduction in symmetry---are required to
produce the observed Mott insulating state. As we have mentioned, a correct calculation
also includes the orbital polarization term, one whose evident effect (from its 
form of the energy)
is to enhance the orbital moment. However, the resulting orbital-dependent potential shifts bands and
the final outcome can be more complicated than a simple enhancement. Finally, all
moments (spin plus orbital) should be treated non-collinearly. A calculation of this
type -- non-collinear GGA+U+OP+SOC -- might be the first of its kind.

\section{Acknowledgments}
We acknowledge many cogent observations from A.~S. Botana throughout the course
of this study, and thank V. Pardo for many discussions on the impact of spin-orbit
coupling and the origin of orbital moments.  This research was supported by
U.S. Department of Energy grant DE-FG02-04ER46111.

\end{document}